\documentclass[prl,a4paper,twocolumn,nofootinbib,preprintnumbers]{revtex4}
\usepackage[]{hyperref}

\usepackage{bbm}
\usepackage{graphicx}
\usepackage{dcolumn}
\usepackage{bm}
\usepackage{textcomp}
\usepackage{epsf}
\usepackage{epsfig}
\usepackage{dcolumn}
\usepackage{float}
\usepackage{epstopdf}
\usepackage{hypcap}
\usepackage{amsmath}

\hypersetup{
  bookmarks=true,         
  unicode=false,          
  pdftoolbar=true,        
 pdfmenubar=true,        
 pdffitwindow=true,     
 pdfstartview={FitH},    
 pdfsubject={Neutrino Oscillations Phenomenology},   
 pdfnewwindow=true,      
 pdfcreator={RevTeX},
 colorlinks=true,       
 linkcolor=red,          
 citecolor=blue,        
filecolor=black,      
 urlcolor=blue,           
  }

\begin{document}

\title{Influence of geometrical perturbations on the ion motional frequencies in a Paul trap }
\author{Manoj Kumar Joshi}
\affiliation{Atomic and Molecular Physics Division, Bhabha Atomic Research Centre, Mumbai
400085, India
}

\author{Pushpa M. Rao}
\email[Email Address: ]{Pushpam@barc.gov.in}
\affiliation{Atomic and Molecular Physics Division, Bhabha Atomic Research Centre, Mumbai
400085, India
}
\begin{abstract}
The effect of geometrical perturbations on ion oscillation frequencies in a $3$-D quadrupole trap is analyzed by theoretical simulations and compared with the experimental results. Theoretically, the distorted potential within this non-ideal trap caused by the insertion of filament has been simulated. An analytical expression for the potential is then fitted and numerical solution of the equations of trapped ion motion is obtained. Motional frequencies are then calculated from the Fourier transformation of the simulated ion trajectories at any given trapping potential and compared with our experimental findings. The shift in the secular frequency with respect to the level of insertion of the filament within the trap is evaluated. 

\end{abstract}
\maketitle


\section{Introduction}
Confinement of charged particles in an ion trap offers unique advantageous allowing state of the art experiments to be performed, resulting in a profound insight in the fields of atomic, nuclear and quantum physics. Ion traps are extensively used in the different fields of scientific applications like mass spectrometry, several atomic and nuclear physics related experiments and more recently in the field of quantum computations \cite{1,2,3,4}. Starting with Paul \cite{5} and Dehmelt \cite{6} several other researchers \cite{7,8,9} have extensively worked in this field presenting innovative and improved models of ion traps. Improvement in design of ion traps and theoretical simulations to achieve the near ideal quadrupole field is an ongoing evolutionary process in the field. Conceptually, trapping of charged particles is not a difficult task but utilization of traps as devices for a quantum engineering, precise mass measurements and high precision studies in atomic and molecular physics, requires an 
appropriate design and thorough understanding of the ion traps. In the past thirty years, an enormous work has been done in the field of ion traps. 

For an ideal $3$-D Paul trap the field inside varies linearly with distance from the trap centre. And potential inside is given as 
\begin{equation}
\label{eq1}
\phi=\frac{U_{dc}+V_{rf} cos(\varOmega t)}{2 r_{0}^{2}} (2z^2-x^2-y^2 ), 
\end{equation}
where $U_{dc}$ and $V_{rf}$ are the amplitudes of the static and dynamic trapping potentials applied across the trap electrodes and $\varOmega$ is the angular frequency of the RF field. The size of trap is defined in terms of radius of ring electrode ($r_0$) and the separation of end-caps ($2z_0$) and the relation $r_0/z_0$ equal to $\surd2$ is maintained.  The ion motion for an ideal trap can be described by Mathieu equations as discussed in \cite{10, 11}. In practice, it is difficult to make an ideal trap that is described in theory. Manufacturing errors and several other modifications made in the trap for the convenience of specific experiments cause deviations in the potential given in equation ($\ref{eq1}$). Any deviation from this `ideal' quadrupole potential is expressed either by multipole expansion of the generalised potential \cite{12} or suitable modifications in the equation ($\ref{eq1}$) \cite{13}. This deviation of the potential from the ideal quadrupole form alters the characteristics of 
the trapped ions, manifesting as nonlinear effects which are broadly classified in three ways.

The first one is where the trapped ions exhibit instabilities even though the trap is operated within the stability region. These instabilities were first seen by Busch and Paul \cite{14} and were termed as black holes and canyons. Measurement of the number of trapped ions at different operating points provides an experimental verification \cite{15,16,17} of the occurrence of these instabilities. Wang et al \cite{18} have given a theoretical explanation for such instabilities wherein they say that at certain operating points the superposition of the motional frequencies and their overtones are an integral multiple of the drive frequency, causing the transfer of energy from the drive field to the trapped ions.

The second effect usually seen is the nonlinear resonance occurring in the resonant excitation and ejection of trapped ions due to coupling of ion motion. This causes the excitation and ejection of trapped ions at frequencies other than the harmonic motional frequencies \cite{19,20,21}. These nonlinear effects are used in isotope separation of ions in a Paul trap \cite{22}.  

The third observed effect in ion traps is the shift in the secular frequencies due to nonlinearities in the trapping field. As ions move away from the centre of a non-ideal trap, where the field within varies nonlinearly with distance, the secular frequencies exhibit a shift from those calculated for an ideal trap. The shift in the ion oscillation frequency depends upon the polarity, strength and order of that particular multipole present \cite{23}. This geometrically driven nonlinear effect exhibits a shift in the stability region and is observed in hybrid \cite{24} and stretched mode \cite{25} of Paul traps. This shift in the stability region is explainable either by numerical methods \cite{26, 27} or by including weak nonlinear potentials in the Mathieu equations that describe the motion of trapped ions \cite{28}.

For an ideal trap the motional frequencies depend solely upon the mass, operating parameters and trap dimensions. The motional frequencies get modified due to the presence of other ions and aberrations in the trap geometry. Several researchers have shown that when ions are collectively excited the motional resonances do not depend on the number of ions trapped ions i.e., no observable shift in the motional resonance frequencies is seen due to the space charge effect \cite{22, 29}. The perturbations in the trapping field due to geometrical aberrations are the main cause of the shift in motional frequencies. These nonlinearities affect high precision studies which rely on the accurate measurements of motional frequencies.   

In almost all cases the potential within a non-ideal trap is considered to be rotationally symmetric. In the current work we discuss the change in the trapped ion characteristics when they are subjected to a slightly asymmetric potential caused due to geometric modifications. In our experiment the filament which is placed slightly above the lower end cap, alters the potential within the trap and also causes the breakage of rotational symmetry. Cumulative effects of the breakage of rotational symmetry and alteration of the trapping potential lead to changes in the trapped ion motion. In this article we mainly discuss the shift in motional frequencies due to the geometry driven perturbations through our experimental and theoretical studies.  
\section{Theory of trapping}
In an ideal Paul trap, the time dependent potential given in equation ($\ref{eq1}$) is responsible for confining the ions close to the trap centre. The motion of trapped ion is described by Mathieu equations \cite{30} and is expressed as
\begin{equation}
\label{eq2}
\frac{d^2 u_i}{d\tau^2}+(a_i - 2q_i cos2\tau) u_i=0 ,
\end{equation}
where $u_i$ represents the position coordinate in Cartesian coordinate system and the normalized time $\tau$ is $\varOmega t/2$. The parameters, $a_i$ and $q_i$ are the functions of trapping potentials and are defined as 
\begin{equation}
\label{az}
a_z = \frac{8QU_{dc}}{m r_0^2 \varOmega^2}, a_x= a_y = \frac{-4QU_{dc}}{m r_0^2 \varOmega^2} ,
\end{equation}
\begin{equation}
\label{qz}
q_z = \frac{-4QV_{rf}}{m r_0^2 \varOmega^2}, q_x= q_y = \frac{2QV_{rf}}{m r_0^2 \varOmega^2} ,
\end{equation}
where $m$ and $Q$ are the mass and charge of trapped species respectively.

The solutions of equation ($\ref{eq2}$) represent the motion of ion in the trap comprised of the secular motion modulated at the drive frequency ($\varOmega$). The secular motional frequency ($\omega_i$) depends on the mass, charge of trapped species and operating parameters of the trap. In the pseudo potential well model \cite{6}, the secular frequency $\omega_i$ is expressed as function of a parameter $\beta_i$, which is a function of $a_i$ and $q_i$ defined in equation ($\ref{az}$) and ($\ref{qz}$). For small values of $a_i$ and $q_i$ ($<0.4$) the parameter $\beta_i$ is approximated to $\sqrt{a_i+q_i^2/2}$. In other cases $\beta_i$ can be written as a solution of a continuous function in $a_i$ and $q_i$ \cite{10, 11}. These expressions are applicable only for an ideal Paul trap and any imperfection calls for modification in the expressions.

\begin{figure}
\begin{center}
\includegraphics[width=0.4\textwidth]{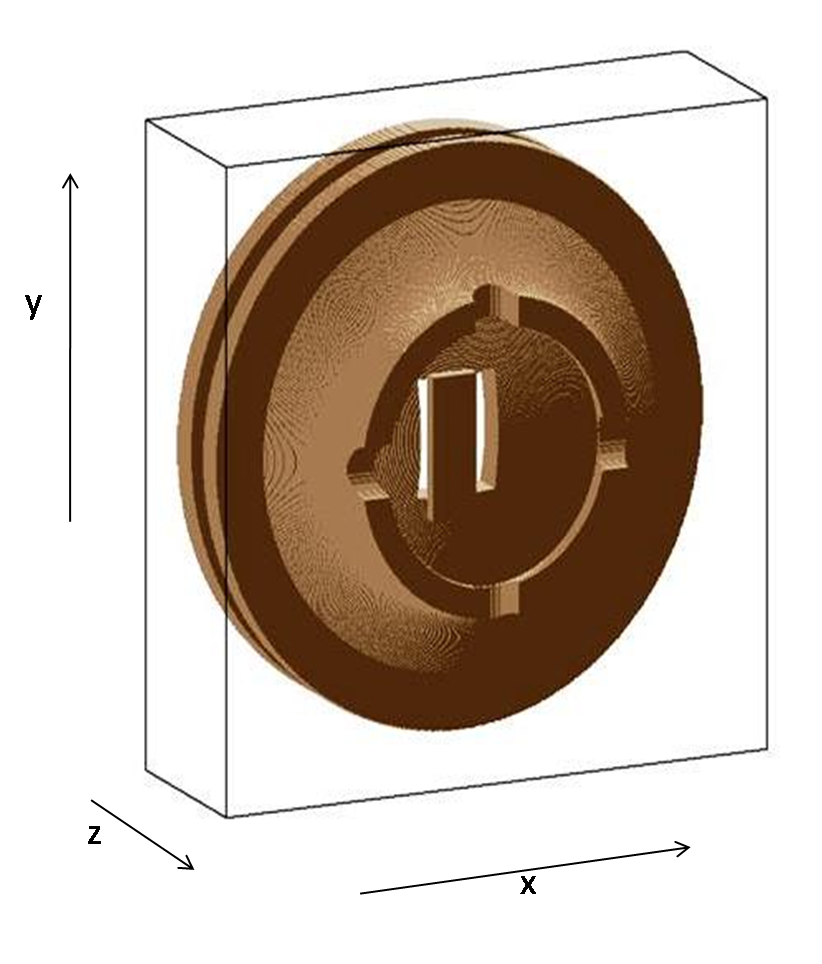}
\caption{ Orthogonal view of the trap geometry with filament raised.}
\label{fig1}
\end{center}
\end{figure}

\section{ Experiment}
The experimental set up used in current studies is similar to that reported in our earlier paper \cite{31}. The ion trap consists of a set of three hyperbolically shaped electrodes. The ring electrode radius ($r_0$) is $20$mm. The ring electrode has four symmetrically positioned holes each measuring $8$mm in diameter to facilitate steering of the laser beam \cite{32}. The lower end-cap has two rectangular slots to place filaments within the trap( figure $\ref{fig1}$). In our experiments a filament of size $22\text{mm}\times 8\text{mm}$ is placed $5$mm above and $6$mm away from the centre of lower end-cap. For trapping, a RF potential at fixed drive frequency ($\varOmega/2\pi = 500$kHz) along with the DC potential was applied to the ring electrode keeping end-caps at the ground potential. The DC and RF potentials were chosen in a way such that the operating parameters ($a_i$, $q_i$) are well within the stability region. Europium ($Eu^+$) ions were loaded in the trap by surface ionization of the sample 
placed on the filament. The base vacuum attained in the trap assembly was $5\times10^{-9}$torr and the ions were continuously loaded in the trap during the experiment.   

\begin{figure}
\centering
\includegraphics[width=0.4\textwidth]{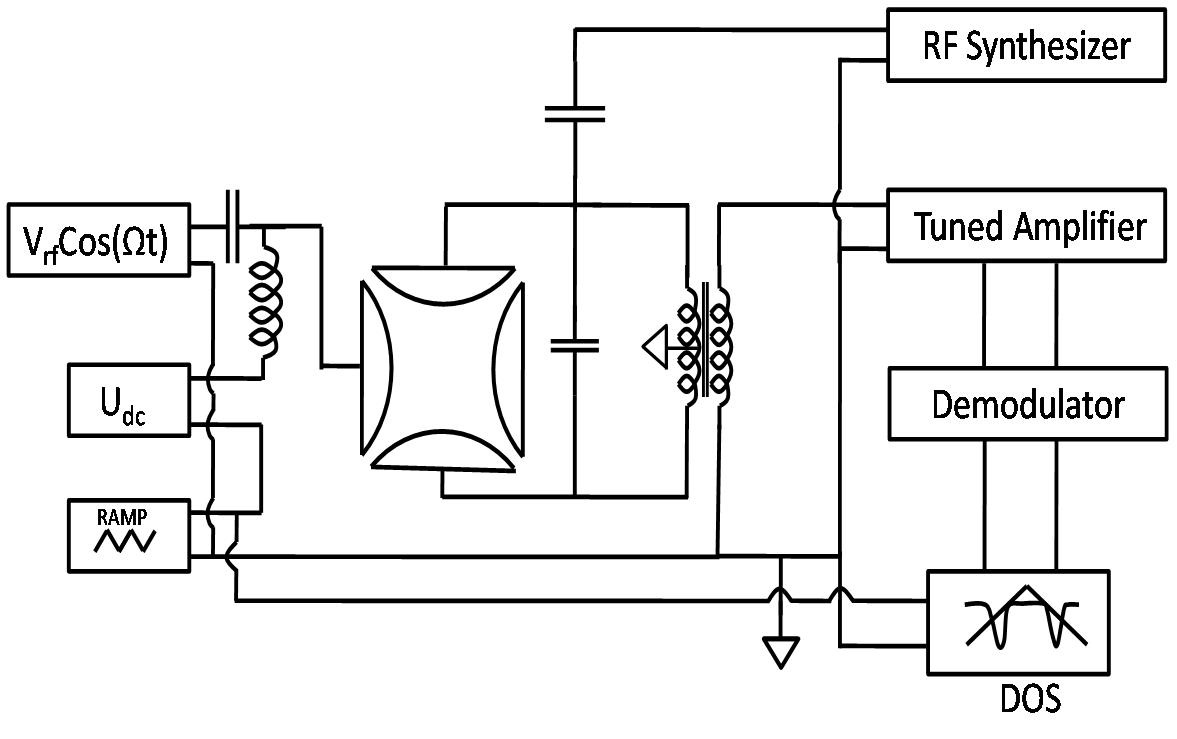}
\caption{ Schematic diagram of the ion trapping setup and detection circuit.}
\label{fig2}
\end{figure}

The trapped ions are non-destructively detected \cite{31} by excitation of the motional frequencies using a weak RF field applied across a high impedance tank circuit ($2$M$\varOmega$) at its resonance frequency. This tank circuit is connected across the end-caps of the trap. Figure ($\ref{fig2}$) shows the schematic of the trapping and detection setup. The ion oscillation frequency is swept across the excitation frequency by varying the DC potential ($U_{dc}$). At resonance, damping of the excitation voltage across the tank circuit takes place and this signal is recorded. The number of ions confined in the trap is estimated from the fitting of the signal obtained to the response signal of an equivalent circuitry of trapped ions \cite{33}. In our experiment the number of ions confined is estimated to be $\sim10^6$.

\section{Theory and simulation}
Using SIMION software \cite{34}, the potential generated within the ion trap used in our experimental studies was theoretically simulated. The 3D potential generated is then fitted to an analytical expression which is a series expansion of the generalized potential considered up to the quadrupole term and is given as 
\begin{equation}
\label{eq5}
\phi = \phi_0\left(\alpha_0+\frac{\alpha_1x+\gamma_1z}{r_0}+\frac{\alpha_2x^2+\beta_2y^2+\gamma_2z^2}{r_0^2}\right),
\end{equation}
where $\phi_0 $ is the potential applied at the ring electrode with the end-caps electrically grounded. Cartesian coordinates $x$, $y$ and $z$ are the position coordinates within the trapping volume considering the trap center as origin. In the above equation the coefficients $\alpha$, $\beta$ and $ \gamma$ represent the strength of the multipolar terms of the potential. In accordance with our experiments, the theoretical simulations have been carried out with the filament kept at the same potential as the end-caps. From figure ($\ref{fig1}$) it is clear that our trap has `$XZ$' plane of symmetry, thus the coefficient for the dipolar term or any higher order odd terms along the $y$-coordinate can be neglected throughout the simulations. Trapped ion trajectories are computed by solving the equations of motion numerically using Runge Kutta $4$ code with the potential given in equation ($\ref{eq5}$). Motional frequencies are then evaluated from the Fourier transformation of the simulated ion trajectories for a 
given set of trapping potentials. We repeat the simulations for different heights of the filament. 

\section{Result and discussion}
\begin{figure}
\centering
\includegraphics[width=0.4\textwidth]{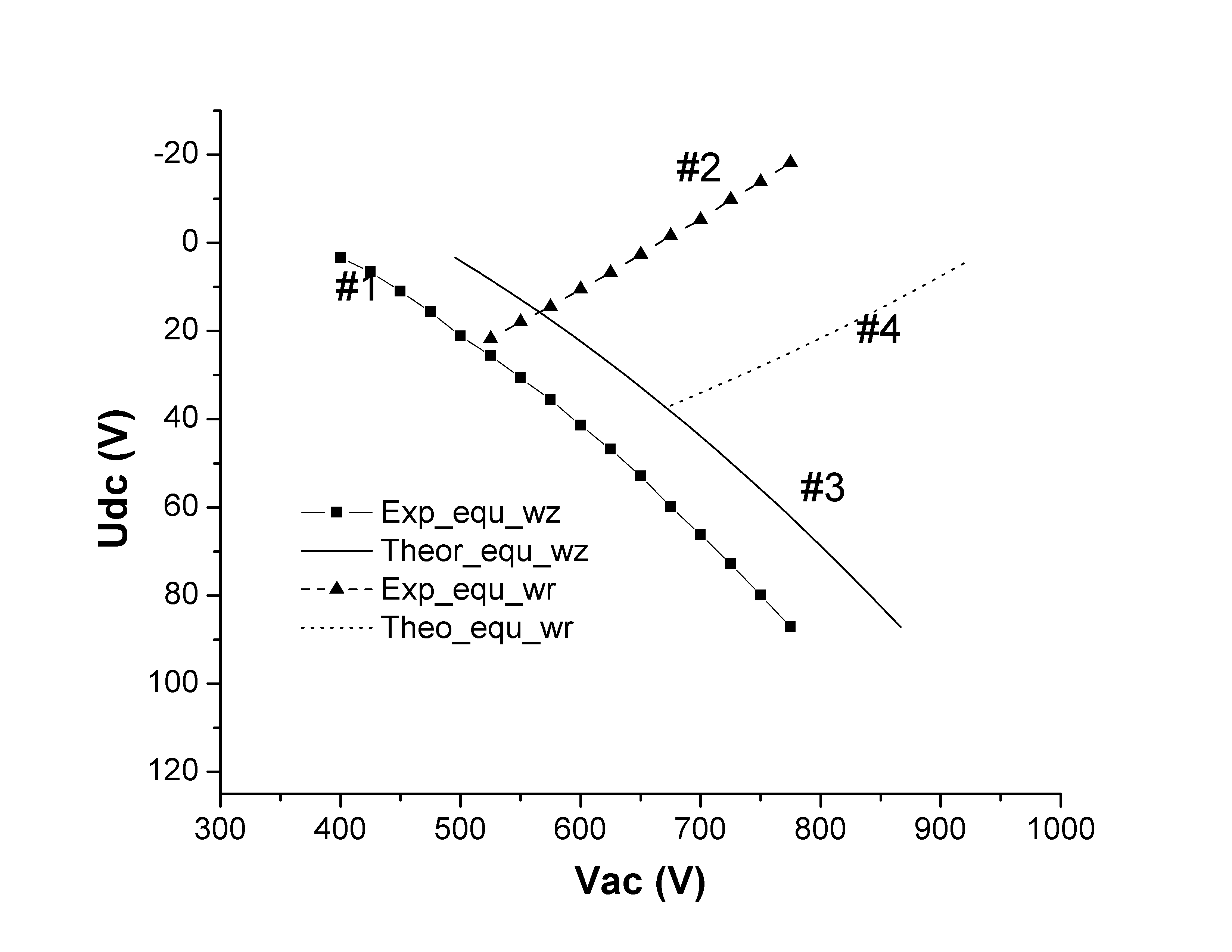}
\caption{ Plot shows theoretical and experimental equi-frequency lines in radial and axial directions. In this plot $\#1$ and $\#2$ are the experimentally observed equi-frequency set of potentials in axial and radial direction respectively. And $\#3$ and $\#4$ are the set of potentials calculated based upon Pseudo potential model for an ideal trap.}
\label{fig3}
\end{figure}

The filament inside the ion trap acts as an extra electrode and causes a deviation in the potential which affects the motional behavior of the trapped ions. To study these effects experimentally, we follow the set of operating potentials ($V_{rf}$ and $U_{dc}$) for which we see the resonant excitation signal at $55.6$kHz. In figure ($\ref{fig3}$) traces $\#1$ and $\#2$ correspond to the experimental set of operating potential that follow the axial and radial equi-frequency lines respectively. For comparison, we evaluated the set of potentials for an ideal trap using pseudo potential model, for which the axial and radial motional frequencies are $55.6$kHz. In figure ($\ref{fig3}$) traces $\#3$ and $\#4$ are the evaluated set of potentials ($V_{rf}$, $U_{dc}$) in the axial and radial modes respectively. From figure ($\ref{fig3}$) it is clear that the sets of operating potentials, at which the damping of resonant excitation signal was observed, differ drastically from the sets of potentials evaluated based on 
the Pseudo potential model. 
This clearly shows that the experimental ion oscillation frequencies differ from those calculated for an ideal trap. We have seen that the calculated ion oscillation frequency for a given set of potential at which resonant excitation signal was seen ($\#1$ and $\#2$ in figure $\ref{fig3}$), differs by approximately $14$kHz from the excitation frequency. This calls for a modification of the expression used to theoretically calculate the ion oscillation frequencies.

For further analysis, we simulated the potential generated within the ion trap using SIMION and fit it to an analytical expression given in equation ($\ref{eq5}$). The coefficients of the least square fit are shown in table $\ref{table}$. Substituting these coefficients in equation ($\ref{eq5}$), the single ion trajectories are simulated for the same set of operating potentials at which we have observed the trapped ion signal. The ion oscillation frequencies both in the radial and axial direction are evaluated from the Fourier transformation of the simulated trajectories.
 
\begin{table}
\centering
\caption{Coefficients calculated from the least square fit.}
\label{table}
\small\addtolength{\tabcolsep}{2pt}
\begin{tabular}{| c | c | c |}
\hline
\textbf{Coefficie-} & \textbf{Calculated} & \textbf{Error} \\ 
\textbf{nts} & \textbf{values} & \textbf{estimated} ($\pm$) \\ 

\hline
$\alpha_0$ & -0.1037 & 0.0008 \\
\hline
$\alpha_1$ & -0.0823 & 0.0012 \\
\hline
$\gamma_1$ & 0.2047 & 0.0014 \\ 
\hline
$\alpha_2$ & 0.6902 & 0.0041  \\ 
\hline
$\beta_2$ & 0.5500 & 0.0041 \\ 
\hline
$\gamma_2$ & -1.2405 & 0.0064 \\ 
\hline
  \end{tabular}
\end{table}

Simulations show a deviation of the potential generated within the trap used in our experiments from an ideal quadrupole potential. It can be seen that the potential within the trap electrodes does not exhibit rotational symmetry. This breakage of rotational symmetry is evident from Table $\ref{table}$, as the coefficients $\alpha_2$ and $\beta_2$ are not equal, $\alpha_1$ and $\gamma_1$ coefficients are not zero. The simulated ion oscillation frequencies in $x$ and in $y$ directions are thus seen to be non equal ($\omega_x \neq \omega_y$). For instance, the simulated motional frequencies for a given set of operating potentials ($V_{rf} = 700$V, $U_{dc}= -5.3$V) are $\omega_x =53.5$kHz; $\omega_y = 41.73$kHz and $\omega_r= 53.5$kHz. 
 
\begin{figure}
\centering
\includegraphics[width=0.4\textwidth]{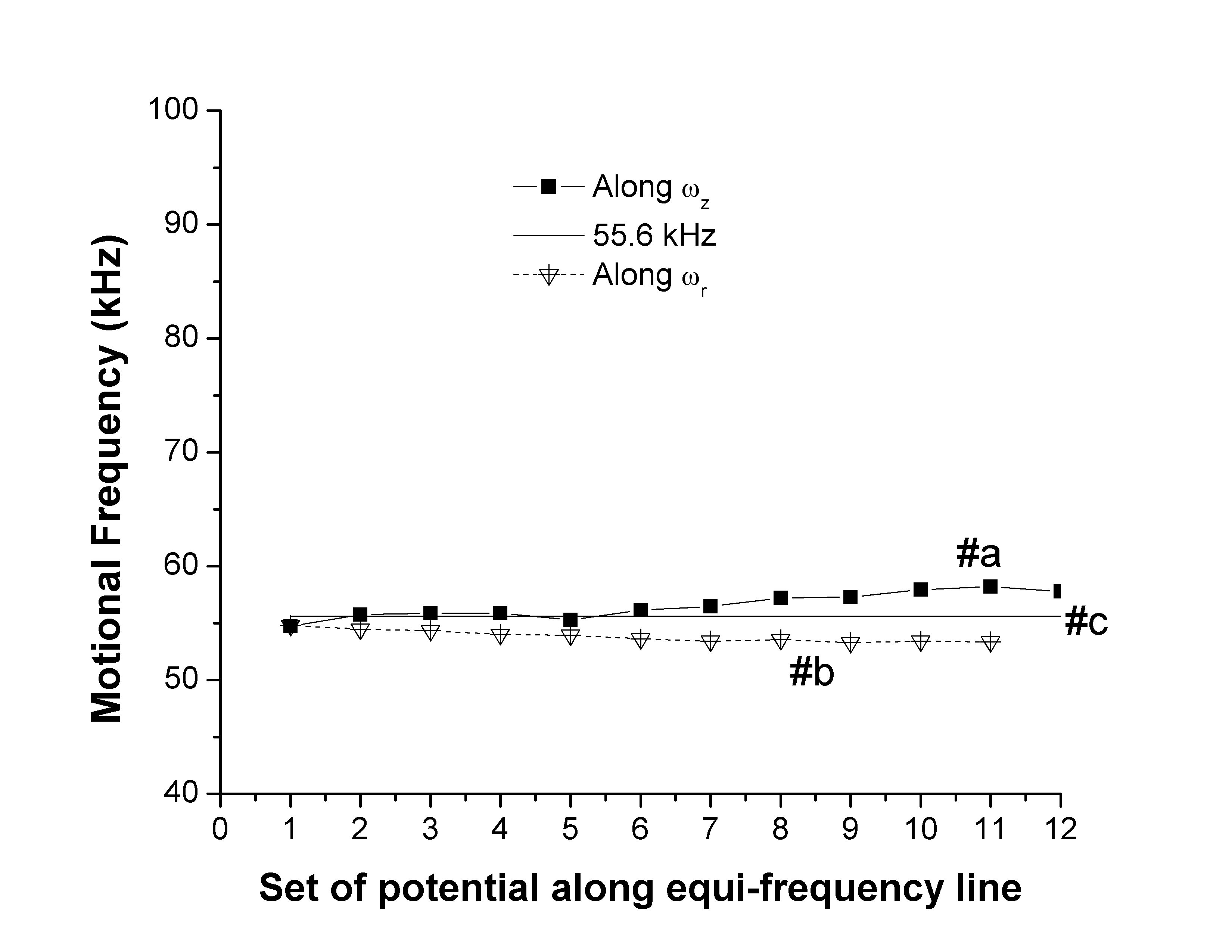}
\caption{ Plot between simulated motional frequencies and operating potentials at which trapped ion signal was observed along the equi-frequency line. In this plot $\#$a and $\#$b are set of potentials that follows the equi-frequency lines in axial and radial direction respectively versus the simulated ion oscillation frequency. And line $\#$c represents the excitation frequency at $55.6$kHz.}
\label{fig4}
\end{figure}

Frequencies in radial ($\omega_r$) and axial ($\omega_z$) directions were simulated at several sets of experimental operating potentials and are compared with the fixed excitation frequency used in the current experiment. In figure ($\ref{fig4}$) the line $\#$c represents the excitation frequency ($55.6$kHz) which is compared with the simulated axial and radial ion oscillation frequencies, shown in lines $\#$a and $\#$b, for the set of operating potentials used in the experiment. In figure ($\ref{fig4}$) each number on the $x-$axis represents a set of operating potentials ($V_{rf}$, $U_{dc}$) in increasing order of $V_{rf}$ values, corresponding to the equi-frequency lines $\#1$ and $\#2$ of figure ($\ref{fig3}$). The $x-$axis in figure ($\ref{fig4}$) maps the set of operating potentials shown along the line $\#1$ and $\#2$ in figure ($\ref{fig3}$) and the simulated frequencies are plotted along the $y-$axis. From figure ($\ref{fig4}$) it is seen that the evaluated axial and radial motional frequencies are 
slightly different from the experimental excitation 
frequency and also exhibit an opposite shift. This can be attributed to the non zero contribution of the higher order multipoles (higher than the quadrupole term) which is not taken into account in the present theoretical simulations.

It is seen that the observed shift in the motional frequencies increases with increasing $q_z$ ($V_{rf}$ ) values and this is attributed to the increase in the kinetic energy of the trapped ions \cite{35}. With increasing kinetic energy the trapped ions move away from the trap centre, leading to a more pronounced effect owing to the contributions of the higher order multipoles. Franzen et al and Sevugarajan et al in their articles \cite{23,28} have theoretically described similar effects on the radial and axial motions due to the higher order multipole contributions (specifically due to the octopole term). 
 
\begin{figure}
\centering
\includegraphics[width=0.4\textwidth]{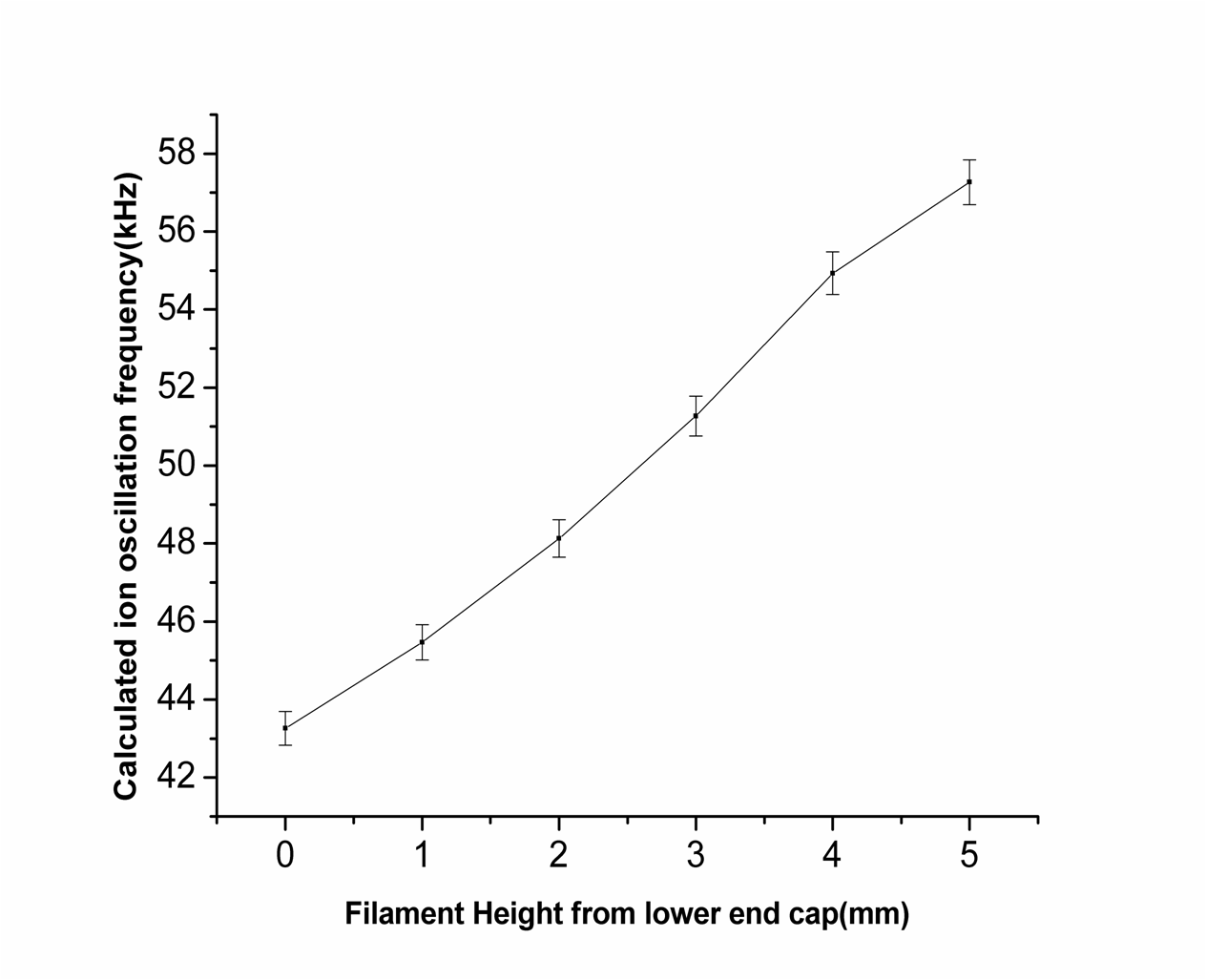}
\caption{ Calculated secular ion oscillation frequency in the axial direction plotted at different filament heights.}
\label{fig5}
\end{figure}

To see the extent of geometrical aberrations and its effect on the ion oscillation frequencies, the axial ion oscillation frequencies were evaluated at different filament heights keeping the operating potential fixed at $V_{rf}=550$V, $U_{dc}=21.75$V. Figure ($\ref{fig5}$) shows a plot of the ion oscillation frequencies simulated at filament heights varied from ${0-6}$mm. From this figure it is clear that the ion oscillation frequency increases as a function of filament height. This is because as the filament is moved towards the trap centre, the changes in the potential are more pronounced.

 \section{Conclusion}
In the present study we have seen the effect of geometrical aberrations in Paul trap on the ion oscillation frequencies. Due to the insertion of filament, the ion trap used is no longer `ideal' and the potential within is also not a pure quadrupole potential. Thus the ion oscillation frequencies evaluated for an ideal trap is no longer valid. The change in the potential shifts the motional frequencies and this has been studied both experimentally and theoretically.
 
Using dipolar excitation, we have been able to resonantly excite and detect the ion oscillations, both in axial and radial directions. This is possible because of the presence of a small quadrupolar contribution in dipolar mode of excitation. Experimental studies wherein equi-frequency lines are traced at several set of operating potentials within the stability region differ drastically from the equi-frequency lines calculated for an ideal quadrupole trap.
 
Theoretically, we have simulated the near exact form of the potential within the trap with geometrical aberrations caused by insertion of the filament. From simulations we see that the coefficients calculated from the multipolar expansion for any given order differs from that of an ideal trap. In addition, a breakage of rotational symmetry is also observed. The motional frequencies evaluated from the simulations for this non-ideal trap is seen to be slightly shifted from the experimental resonant excitation frequency. Further, the ion oscillation frequencies have been simulated for different filament heights. We see that the simulated axial ion oscillation frequency increases as the filament is moved towards the trap centre.

\bibliographystyle{apsrev}
\bibliography{manoj}
\end{document}